\documentclass[usenatbib]{mn2e}
\pdfoutput=1

\usepackage{newtxtext,newtxmath}
\usepackage[T1]{fontenc}
\usepackage{ae,aecompl}
\usepackage{docmute}
\usepackage{float}
\usepackage{comment}
\usepackage{color}
\usepackage{caption}
\usepackage{multirow}
\usepackage{docmute}
\usepackage{import}
\usepackage{graphicx}	
\usepackage{amsmath}	
\usepackage{amsfonts}
\usepackage{booktabs}
\usepackage{siunitx}
\usepackage[dvipsnames]{xcolor}
\usepackage{fixltx2e}
\usepackage{siunitx}
\usepackage{pifont}
\usepackage{hyperref}
\usepackage{aas_macros}
\hypersetup{linkcolor=black,citecolor=black,urlcolor=magenta,colorlinks=true}

\title[Pop III IMF from transients]{
Probing the initial mass function of the first stars with transients}
\vspace{-0.5cm}

\vspace{-0.2cm}
\author[Lazar \& Bromm]{
Alexandres Lazar$^{1}$\thanks{\href{mailto:aalazar@uci.edu}{aalazar@uci.edu}}
and Volker Bromm$^{2}$
\\
$^{1}$Department of Physics and Astronomy, University of California, Irvine, CA 92697, USA\\
$^{2}$Department of Astronomy, The University of Texas at Austin, Austin, TX 78712, USA
\vspace{-0.4cm}
}

\date{\vspace{-0.6cm}}
\pubyear{2021}

\begin{document}
\label{firstpage}
\pagerange{\pageref{firstpage}--\pageref{lastpage}}
\maketitle

\vspace{-0.2cm}
\begin{abstract}
The emergence of the ﬁrst, so-called Population III (Pop III), stars shaped early cosmic history in ways that crucially depends on their initial mass function (IMF). However, because of the absence of direct observational constraints, the detailed IMF remains elusive. Nevertheless, numerical simulations agree in broad terms that the ﬁrst stars were typically massive and should often end their lives in violent, explosive deaths. These fates include extremely luminous pair-instability supernovae (PISNe) and bright gamma-ray bursts (GRBs), the latter arising from the collapse of rapidly rotating progenitor stars into black holes. These high-redshift transients are expected to be within the detection limits of upcoming space telescope missions, allowing to place eﬀective constraints on the shape of the primordial IMF that is not easily accessible with other probes. This paper presents a framework to probe the Pop III IMF, utilizing the cosmological source densities of high-redshift PISNe and GRBs. Considering these transients separately could provide useful constraints on the Pop III IMF, but tighter bounds are obtainable by combining PISN and GRB counts. This combined diagnostic is more robust as it is independent of the underlying Pop III star formation rate density, an unknown prior. Future surveys promise to capture most high-redshift GRBs across the entire sky, but high-redshift PISN searches with future telescopes, e.g. Roman Space Telescope, will likely be substantially incomplete. Nevertheless, we demonstrate that even such lower bounds on the PISN count will be able to provide key constraints on the primordial IMF, in particular, if it is top-heavy or not.
\end{abstract}


\raggedbottom
\section{Introduction}
One of the key goals in modern astrophysics and cosmology is to understand the formation of the first structures in the Universe, and how they shaped early cosmic evolution \citep[e.g.][]{Bromm2011,Loeb2013,Dayal2018}. Among those objects are the elusive first generation of metal-free stars, the so-called Population III (Pop III), predicted to have formed at redshifts $z \simeq 10-30$ \citep{Abel2002,Bromm2004}. The emergence of the first stars marks the final moments of the cosmic dark ages, when the simple conditions of the early Universe were transformed into a state of increasing complexity, owing to the production of ionizing photons and the initial enrichment with heavy chemical elements during the first billion years after the Big Bang. 

An essential ingredient for all models of Pop III star formation is the initial mass function (IMF), as it largely shapes the impact of the first stars on early cosmic history. At present, however, the exact form of the primordial IMF remains uncertain (see \citealt{Chabrier2003} for a comprehensive review), given that Pop III stars have not yet been directly observed. Even though the upcoming {\em James Webb Space Telescope} ({\em JWST}) will be able to probe the early Universe out to $z\approx 15$, it will be unable to observe the very first stars, due to the low luminosities of their host haloes \citep{Zackrisson_2015,Jeon2019}. The direct detection of Pop III may thus have to await future facilities, such as the ``Ultimately Large Telescope'' (ULT), a 100m liquid-mirror design to be constructed on the Moon \citep{Schauer2020}. In principle, any detection prospect could be boosted through strong gravitational lensing  \citep{Zackrisson2012,Rydberg2015,Wong2019,Vikaeus2021}, but meaningful constraints for the primordial IMF would require a large sample of lensing events that will be unattainable in the foreseeable future.

Moreover, even with significant progress made in multi-physics, high-resolution simulations \citep[e.g.][]{Hosokawa2016,Stacy2016,Hirano2017,Sharda2019}, converging on a robust prediction of the detailed IMF will likely be out of reach for such a first principles theoretical approach. Therefore, obtaining guidance from observations other than the direct detection of UV stellar light is of great importance to constrain the characteristics of the primordial IMF. 

One of the most promising routes of indirectly probing the IMF of the first stars is through {\em stellar archaeology} --- the approach of surveying local stellar fossils for vestiges of the early Universe --- based on observing the elemental abundance patterns in extremely metal poor (EMP) stars. In principle, this would allow us to distinguish different supernova (SN) types that have enriched the gas out of which second-generation, Population~II (Pop II), stars form \citep{Beers2005,Frebel2010,Chen2017a,Ishigaki2018,Hartwig2019}. The nucleosynthetic imprint on Pop II stars would thus impose bounds on the IMF of the first stars, with numerous studies along these lines \citep[e.g.][]{Tumlinson2006,Salvadori2010,Kulkarni2013,Bennassuti2014,Hartwig2015}. The potential power of this approach is expected to be fully harnessed with future large, high-resolution spectroscopic surveys \citep[e.g.][]{Jeon2021}. In particular, local ultra-faint dwarf (UFD) galaxies may provide ideal laboratories to probe the first stars \citep[e.g.][]{Jeon2017,Rossi2021}, with a complete census of the UFD stellar abundances within reach of the next-generation of extremely large telescopes.

A different strategy is to probe individual Pop III stars at the moment of their death, predicted to involve violent, hyper-energetic events, with rates that depend on the underlying IMF \citep[e.g.][]{Bromm2013b}. One such fate is a pair-instability SN
\citep[PISN;][]{HegerWoosley2002,Mackey2003}, predicted for progenitors within the mass range of $140-260\ M_{\odot}$ \citep{HegerWoosley2010}, possibly extending down to $\sim 85\ M_{\odot}$ in case of rapid rotation \citep{Chatzopoulos2012,Yoon2012}. Typically, Pop III stars are expected to end their lives as core-collapse SNe (CCSNe) for progenitors in the mass range of $8-40\, M_{\odot}$, or as energetic hypernovae for high rotation rates \citep{Umeda2005}. 
Observations of Pop III SNe could thus directly probe the Pop III IMF since, in principle, they can be detected at high redshifts, and are distinguishable in terms of progenitors of different mass. In particular, PISNe are exceptionally luminous, compared to CCSNe, rendering them excellent beacons of the high-redshift Universe
\citep{Tanaka2012,Whalen2013a,Whalen2013c,Souza2014,Hartwig2018b,Moriya2019,Rydberg2020}.

Pop III stars, given their high characteristic mass, are also predicted to give rise to highly energetic gamma-ray bursts (GRBs), with prompt emission and afterglow components \citep{BrommLoeb2006,Belczynski2007,Toma2016,Burlon2016,Kunigawa2019}. Importantly, GRBs are the most powerful electromagnetic explosions in the Universe and are detectable across a wide range of wavelengths up to high redshifts \citep[e.g.][]{Ciardi2000,Lamb2000,Gou2004,Meszaros2010,Ghirlanda2015}. The prospect of high-redshift GRBs is demonstrated by the spectroscopic detection of GRBs out to $z\sim 8.2$ \citep{Salvaterra2009,Tanvir2009}, and photometrically to $z\sim9.4$ \citep{Cucchiara2011}. Recently, \cite{Jiang2021} attributed a near-infrared transient to the $z\sim 11$ galaxy GN-z11 \citep{Oesch2016}, arguing that it originates in a long GRB within this extremely distant host \citep{Kann2020}. However, the GRB interpretation is debated, with alternative explanations invoking reflections from near-Earth objects \protect\citep{Nir2021,Michalowski2021}. Formation of the central engine for long-duration GRBs is often explained by the collapsar model \citep{Macfadyen2001}, which invokes the direct collapse of a rapidly-rotating, massive star into a black hole (BH). This association with the death of massive stars naturally explains the observed location of long-duration GRBs in star-forming regions \citep{Blain2000,Langer2006,Fruchter2006}.

The next generation of telescopes promises to extend the frontier of observational cosmology to even higher redshifts. While it is unlikely that the first stars themselves will be visible, even with the {\em JWST}, the source density of energetic transients produced by Pop~III stars could be surveyed over a wide field of view (FOV) at near-infrared (NIR) wavebands.
The upcoming {\em Roman Space Telescope}\footnote{Formerly known as Wide-Field InfraRed Survey Telescope (WFIRST).} ({\em RST}; \citealt{Spergel2015}) could be effective in the search for Pop III PISNe out to redshifts of at least $z \sim 6$, given its wide-field NIR instrument with a 10 deg$^{2}${} FOV \citep{Whalen2013c,Moriya2019}.
In addition, GRB astronomy has made great strides, in particular since the launch of the {\em Swift} satellite with its large sample of redshift determinations \citep[see][]{Gehrels2009}. Future missions currently under development target the still elusive highest redshift events, such as the High-$z$ Gamma-ray bursts for Unraveling the Dark Ages Mission (HiZ-GUNDAM; \citealt{Yonetoku2014,Yoshida2016}), the Transient High-Energy Sky and Early Universe Surveyor (THESEUS; \citealt{Amati2021,Tanvir2021}), and the Gamow Explorer \citep{White2020}. These missions promise to probe GRBs out to the highest redshifts, thus pushing the cosmic horizon to the very beginning of star and galaxy formation \citep[see][]{Toma2016}.

In this paper, we explore the possible constraints on the Pop III IMF from high-redshift transients, specifically targeting GRBs and PISNe triggered by the first stars. In Section~\ref{sec:model}, we present the foundations for our analysis, the cosmological number density of sources, the generalized Pop III IMF, and the GRB and PISN formation physics. Based on this framework, in Section~\ref{sec:observability} we discuss how the relative occurrence of those transients can constrain the key IMF parameters, how future missions and surveys can carry out this determination, and possible caveats with our methodology. We offer our concluding remarks in Section~\ref{sec:conclusions}.

\section{Modeling the transient event rate}
\label{sec:model} 
In this section, we summarize our basic expressions for the cosmological rate of Pop III transients and a generalized form of the IMF, as well as our idealized assumptions on how PISN and GRB deaths are connected to Pop III star formation. Throughout this paper, we
assume a standard flat $\Lambda$CDM cosmology, where $\Omega_{\rm m} = 1 - \Omega_{\Lambda}=0.3$ and $h=0.7$, where $h$ is the dimensionless Hubble constant, $H_{0} = 100h\ \rm km\ s^{-1}\ Mpc^{-1}$.

\subsection{Observable number of transients}
In the Earth's reference frame, the rate of observable transient events out to a given redshift $z$, per unit solid angle, $\Omega_{\rm obs}$, is given by \citep[e.g.][]{Mackey2003,Hummel2012}
\begin{align}
    \frac{\mathrm{d}^{2}N}{\mathrm{d}{t_{\rm obs}}\mathrm{d}{\Omega_{\rm obs}}}(<z)
    &=
    \eta_{\rm IMF}
    \int_{0}^{z}\frac{\mathrm{d}z'}{(1+z')} 
    \frac{\mathrm{d}^{2}V}{\mathrm{d}z'\, \mathrm{d}\Omega_{\rm obs}}
     \,  \frac{\mathrm{d}^{2}\mathcal{M}_{*}}{\mathrm{d}t_{\rm em}\mathrm{d}V}(z')
    \, ,
    \label{eq:fov.times}
\end{align}
which is determined by the comoving star formation rate density (SFRD), $\mathrm{d}^{2}\mathcal{M}_{*}/(\mathrm{d}t_{\rm em}\mathrm{d}V)$. Here, $\eta_{\rm IMF}$ accounts for the number of sources found within the appropriate range of progenitor masses for a given final fate (in our case, death as a PISN or GRB), per unit stellar mass. This fraction is governed by the underlying IMF, in particular its degree of top-heaviness (see below). Note that the cosmological time dilation is accounted for by the $(1+z)$ factor, when relating time in the observer frame, $t_{\rm obs}$, to that measuring emission at the source, $t_{\rm em}$. The differential comoving volume element can be written as 
\begin{align}
    \frac{\mathrm{d}^{2}V}{\mathrm{d}z\, \mathrm{d}\Omega_{\rm obs}}(z) &= r^{2}(z) \left| \frac{\mathrm{d}r}{\mathrm{d}z}\right| \, ,
\end{align}
where $r(z)$ is the comoving distance to redshift $z$,  
\begin{align}
    r(z) &= \frac{c}{H_{0}}\int_{0}^{z}
    \frac{\mathrm{d}z'}{E(z')}
    \, .
\end{align}
As usual, $c/H_{0}$ is the Hubble distance, and
\begin{align}
    E(z) &= 
    \sqrt{\Omega_{\rm m}(1+z)^{3} + \Omega_{\Lambda}}
\end{align}
for our flat cosmology.

\raggedbottom
\subsection{Generalized form of the primordial IMF}
\label{sec:imf}
In exploring the IMF for the first stars, we will assume a generalized form by expressing the number of stars per unit stellar mass $m$ \citep{Chabrier2003}:\footnote{Note that definitions of the IMF vary, such that the power-law exponent is expressed with either $x$ or $\alpha$, based on the mass bin spacing, such that $\xi(\log{m}) \propto m^{-x}$ and $\xi(m) \propto m^{-\alpha}$, respectively, where $\alpha = x+1$.}
\begin{align}
    \xi(m) := \frac{\mathrm{d}N}{\mathrm{d}m}
    \propto 
    m^{-\alpha} \exp\Bigg[ - \Big(\frac{m_{\rm char}}{m}\Big)^{\beta} \Bigg] 
    \label{eq:imf}
    \, ,
\end{align}
which approaches $\xi(m) \propto m^{-\alpha}$ at large $m$, and is exponentially suppressed below the characteristic mass, $m_{\rm char}$, as moderated by $\beta$. Throughout the literature, the stellar IMF is discussed in terms of numerous functional forms \citep[see][]{Bastian2010}, but our generalized expression can accommodate most of them. A convenient way to characterize the nature of the IMF, in particular its degree of top-heaviness, is by the average stellar mass, 
\begin{align}
    \bar{m} \equiv \frac{\int \mathrm{d}m\,m\, \xi(m)}{\int \mathrm{d}m\,  \xi(m)}
    \, ,
\end{align}
which is similar in value to $m_{\rm char}$. For example, assuming a \cite{Salpeter1955} slope of $\alpha=2.35$, and with $\beta=1.6$, we have $\bar{m} \approx 2\, m_{\rm char}$. 

For a given IMF, we can evaluate in Equation~\ref{eq:fov.times} the formation efficiency $\eta_{\rm IMF}$, defined as the number of progenitors per unit stellar mass that will lead to one of the final fates considered here, as follows:
\begin{align}
    \eta_{\rm IMF} \equiv 
    \frac{\int_{X} \mathrm{d}m\, \xi(m)}{\int \mathrm{d}m\, m\, \xi(m)}
    \, ,
    \label{eq:imf.efficiency}
\end{align}
where the integral $\int_{X}$ is over the lower and upper (initial) mass limits of the progenitor stars that end their lives in a given way (CCSN, PISN, or black hole).

In our subsequent analysis, we broadly explore the Pop III IMF with a slope ranging from $\alpha \in [-1,3]$ and with characteristic masses of $m_{\rm char} \in [10-100]\times M_{\odot}$. We further assume that $\beta=1.6$ \citep[e.g.][]{Wise2012,Jeon2015} for our analysis. For the remainder of this paper, when we refer to the `IMF', we mean the Pop III IMF. For comparison, the Pop II IMF is assumed to closely follow the observed Pop I IMF \citep[see][]{Chabrier2003}, with typical parameters: $\alpha\simeq 2.35$ and $m_{\rm char} \simeq 1 M_{\odot}$. We expect Pop III stars to extend over a mass range of $m\in [1,300]\times M_{\odot}$, compared with a typical range for metal-enriched populations of $[1,100]\times M_{\odot}$. 

\begin{figure}
    \centering
    \includegraphics[width=1.0\columnwidth]{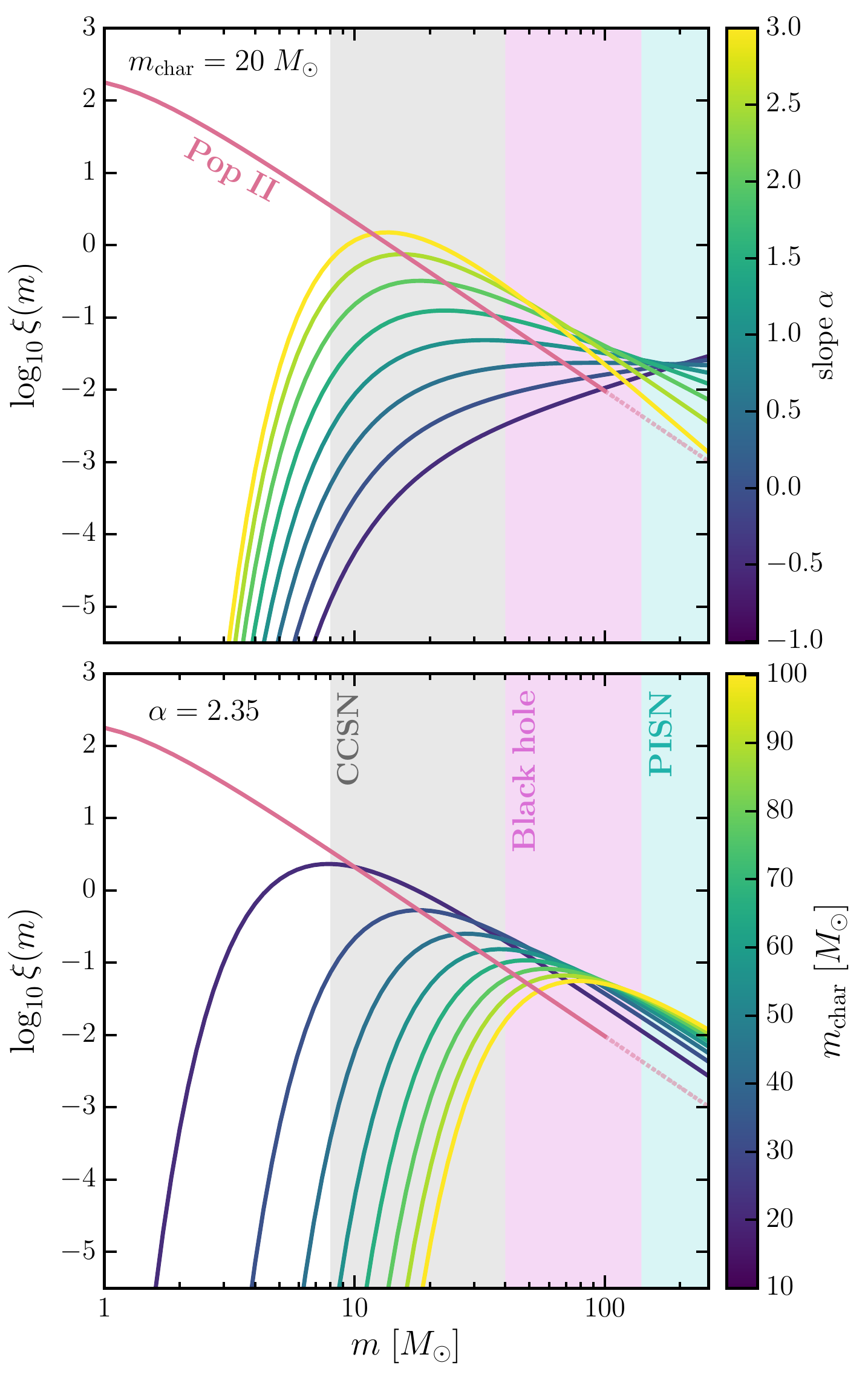}
    \caption{
    Illustrative examples of the Pop III IMF (see Section~2.2), for a fixed $m_{\rm char}=20\ M_{\odot}$ with different values of the slope (top), and for fixed $\alpha=2.35$ with different values of $m_{\rm char}$ (bottom), normalized to the same total stellar mass ($M_{\star}=10^{3}\, M_{\odot}$). We also show the Pop II IMF, assuming $\alpha = 2.35$ and $m_{\rm char}=1\, M_{\odot}$ in Equation~\ref{eq:imf}, and tapering off for $m>100\ M_{\odot}$ (red curve). Each panel indicates the range for the different stellar fates, i.e., CCSNe, PISNe, and black holes \protect\citep{HegerWoosley2002}, allowing us to see how their occurrence depends on the IMF, based on the values of $\alpha$ and $m_{\rm char}$.
    }
    \label{fig:1}
\end{figure}

Figure~\ref{fig:1} provides illustrative examples of the IMF for fixed $m_{\rm char}$, varying the slope (top plot), and varying $m_{\rm char}$ for a fixed slope $\alpha$ (bottom plot). In each case, the parameter variation is indicated with the respective color bar. For comparison, we plot the Pop II IMF (red curve), over its appropriate initial mass range. All IMFs in Figure~\ref{fig:1} are normalized to the same total stellar mass: $M_{\star} = 10^{3}\ M_{\odot}$. The colored bands depict the range of (initial) masses for the formation of CCSNe (gray), black holes (pink), and PISNe (cyan). Here, we focus on the contribution of a given IMF to the last two final fates (pink and cyan bands).

\subsection{Formation of stellar transients}
The extremely energetic PISN and GRB events in the early Universe may be sufficiently bright to allow direct detection. To track their formation, based on the formalism presented above, we select parameters that are motivated by our current theoretical framework for how Pop III stars evolve towards these two end points. In so doing, we emphasize the uncertainties that derive from our limited understanding of the complex physics involved.

\subsubsection{Star-formation rate density}
For our modeling of the SFRDs as a function of redshift, we utilize the convenient power-law form from \cite{Madau2014}:
\begin{align}
    \frac{\dot{\rho}_{\star}(z)}{M_{\odot}\, \rm yr^{-1}\, Mpc^{-3}}
    &= \frac{a (1+z)^{b}}{1 + [(1+z)/c]^{d}} \, .
    \label{eq:SFRD.powerlaw}
\end{align}
The observed Pop I/II SFRD is expressed with best-fitting parameters of: $a = 0.015$, $b = 2.7$, $c=2.9$, and $d = 5.6$ \citep{Madau2014}. As for Pop III, \cite{LiuBromm2020} carried out cosmological simulations to derive the corresponding SFRD, and provided best-fit parameters of its redshift evolution using Equation~\ref{eq:SFRD.powerlaw}. At $z=10$, their Pop III SFRD peaks at $\sim 10^{-4}\, M_{\odot}\rm yr^{-1}Mpc^{-3}$. We here instead normalize to a larger peak value of $\sim 10^{-3}\ M_{\odot}\rm yr^{-1}Mpc^{-3}$ in order to be consistent with the \cite{Jaacks2019} results, but we otherwise assume that the SFRD evolves in the same way as in \citet{LiuBromm2020}. The corresponding parameters used in this paper are $a = 7657$, $b = -5.92$, $c=12.83$, and $d = -8.55$, and the resulting Pop III SFRD is broadly consistent with previous numerical estimates. We acknowledge the considerable uncertainties here, in particular given the absence of any direct observations of Pop III star formation. As we will see later in Section~\ref{sec:combined}, the IMF can be probed independently of any assumed Pop~III SFRD model.

\subsubsection{Assumptions on stellar mass loss}
\label{sec:mass.lost}
Mass loss is a crucial parameter in any model of stellar evolution.
Pop III stars, by definition, are not enriched with metals. Therefore, line-driven stellar winds are expected to be negligible, as the effects of radiation pressure are insignificant for a Pop III star \citep{Kudritzki2000,Baraffe2001,Vink2001,Krticka2006,Ekstrom2008}. Note that massive Pop III stars might become unstable to the pulsational instability, potentially causing mass loss \citep{Baraffe2001}. Another possibility would apply for the case of rapid rotation, driving the surface layer towards the $\Omega\Gamma$-limit \citep[e.g.][]{LiuWind2021,Murphy2021}. However, we will not consider such effects here. For simplicity, we assume that the final masses of Pop III stars are identical to their initial masses, based on our current understanding of the radiation-hydrodynamics of winds. For metal-enriched (Pop I/II) stars, mass loss is inevitably stronger due to radiation-driven winds. In this paper, we focus on the occurrence of Pop III transients, and will therefore not discuss the complex physics of stellar mass loss further.  

\subsubsection{Pair-instability supernovae}
To undergo a PISN event, non-rotating Pop III stars are predicted to have progenitor masses of $m \in [140,260]\times M_{\odot}$ \citep{HegerWoosley2002,Schneider2004,Dessert2013,Takahashi2016}. For Pop III stars, the PISN formation efficiency per unit stellar mass can be computed for a given IMF via Equation~\ref{eq:imf.efficiency}, using this range in masses. For the
combination of $\alpha$ and $m_{\rm char}$ values considered here, we find efficiencies within the range of $\eta_{\rm PISN} \approx [0.2-4]\times10^{-3}\, M_{\odot}^{-1}$. Numerically, this is comparable to the efficiency of (Type~II) CCSNe, originating from Pop I/II stars, where ${\sim} 1/136\, M_{\odot}^{-1} \approx 7\times 10^{-3}\, M_{\odot}^{-1}$ \citep{Bromm2013a}. As mentioned above, we assume that massive Pop I/II stars will lose a significant amount of mass. Thus, even if such stars were born with the large masses required for a PISN, they would rapidly be driven to lower masses. We therefore assume that {\it only} Pop III stars could give rise to a PISN event (see below).

\begin{figure*}
    \centering
    \includegraphics[width=0.975\textwidth]{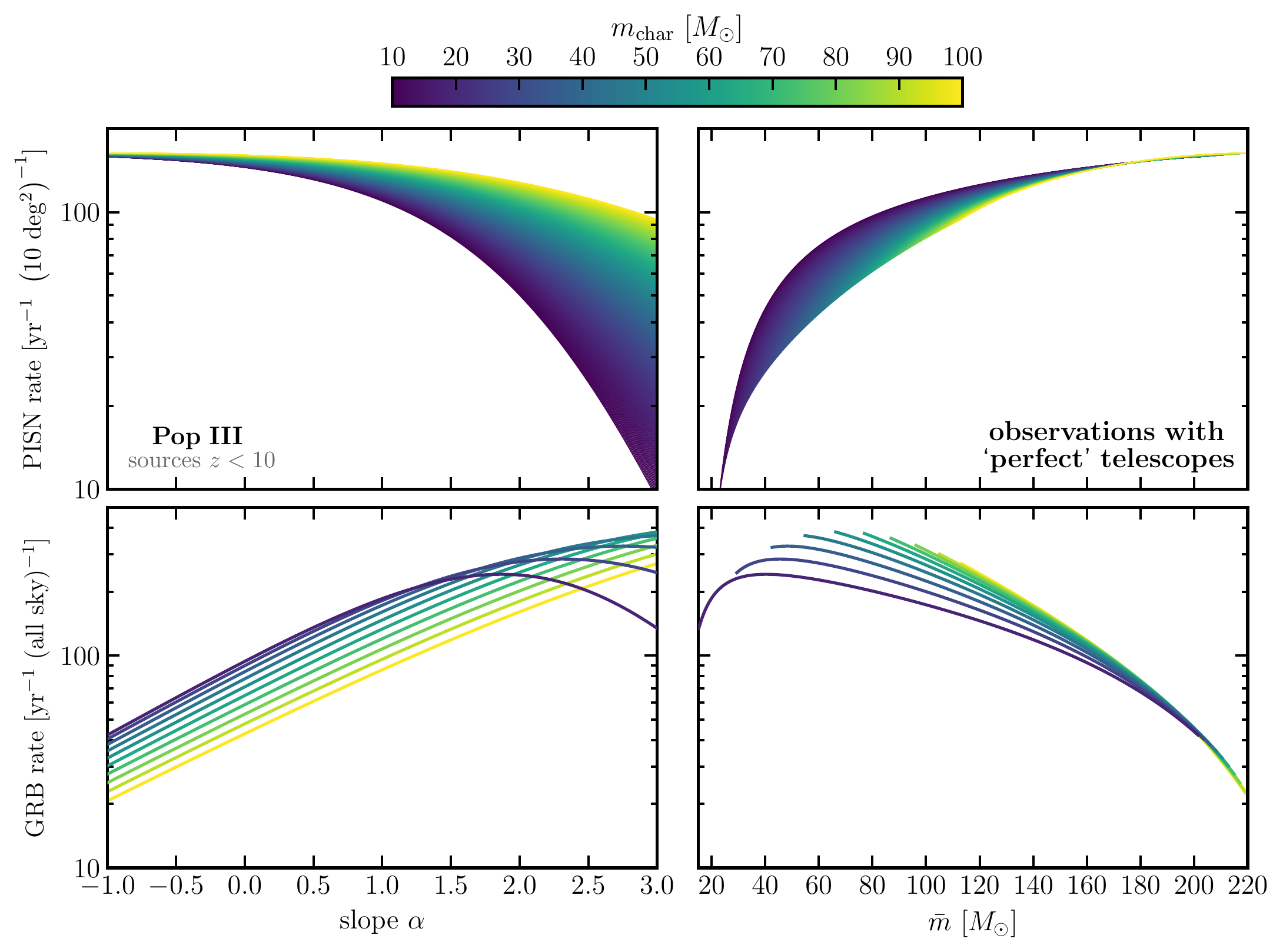}
    \caption{
    Maximal rate of select Pop~III transients. We here assume the idealized case of a `perfect' telescope, without imposing any flux sensitivity limit. {\it Top row:} Pop~III PISN rates, per 10 deg$^2$, as a function of the IMF parameters. {\it Bottom row:} Corresponding all-sky GRB rates. In each case, we consider transient events out to $z = 10$ in the observer frame. {\it Left column:} Dependence on the IMF slope ($\alpha$). {\it Right column:} Dependence on the average stellar mass ($ \bar{m} $). Both are mapped to $m_{\rm char}$, as indicated by the color-bar. It is evident that Pop~III transients can be quite rare, even if considered in the context of perfect observational facilities. 
    }
    \label{fig:2}
\end{figure*}

\subsubsection{Gamma-ray bursts}
\label{sec:fromation.GRB}
At the end of their short lives, it is likely that at least a fraction of the first stars experience the collapse of a rapidly rotating core into a BH, giving rise to a collapsar. Assuming that the formation of long-soft GRBs closely follows the cosmic SF history with no significant time delay \citep[e.g.][]{Conselice2005}, we can estimate their formation efficiency per unit stellar mass, $\eta_{\rm GRB}$, with the formalism above, linking the transient rate to the underlying SFRD. While the detailed behaviour of $\eta_{\rm GRB}$ is uncertain, such as any dependence on redshift and environment, we assume for simplicity that it is proportional to the BH formation efficiency \citep[e.g.][]{BrommLoeb2006}: 
\begin{align}
    \eta_{\rm GRB} &= f_{\rm GRB}\, \eta_{\rm BH}
    \, .
    \label{eq:GRB.efficiency}
\end{align}
Here, $f_{\rm GRB}$ is the fraction of BHs that trigger a GRB, and $\eta_{\rm BH}$ is computed from Equation~\ref{eq:imf.efficiency} for an initial mass range of $m\in[40,140 ]\times M_{\odot}$, for non-rotating Pop III stars \citep{HegerWoosley2002}. Stars with initial masses of $m>260\ M_{\odot}$ could also directly collapse into BHs, due to the endothermic photo-disintegration of heavy nuclei by ultra-energetic gamma-ray photons, sapping the dying star of its (radiation) pressure support. 

The GRB fraction for Pop III is assigned based on the results of high-resolution numerical simulations of the primordial star formation process, providing constraints on the key properties of Pop III stars \citep[reviewed in][]{Bromm2013b}: 

\begin{align}
    f_{\rm GRB} \simeq f_{\rm B}\, f_{\rm close}\, f_{\rm beam}\, f_{\rm rot}\, ,
\end{align}
where $f_{\rm B} \simeq 0.4$ is the binary fraction \citep{Stacy2013,LiuBinary2021}, $f_{\rm close}\sim 0.4$ the fraction of binaries that are sufficiently close to undergo Roche-lobe overflow \citep{BrommLoeb2002}, and $f_{\rm beam}\simeq 1/50 - 1/500$ the beaming factor \citep{BrommLoeb2006}. The first stars are assumed to be rapidly rotating \citep{Stacy2011,HiranoB2018}, such that $f_{\rm rot}\simeq 1$. Here we consider a lower bound for the beaming factor of $f_{\rm beam} \simeq 1/500$. Combining these ingredients, we derive efficiency factors of $\eta_{\rm GRB} \approx [0.1-2]\times10^{-6}\, M_{\odot}^{-1}$ for the range of $\alpha$ and $m_{\rm char}$ values considered in our analysis. These estimates are in good agreement with the study of \cite{Souza2011}, which found $\eta_{\rm GRB}\sim 10^{-6}\, M_{\odot}$.

For comparison, we also briefly discuss the corresponding efficiency for Pop~I/II GRBs, and we note that we do not try to further distinguish any specific contribution provided by Pop~II (metal-poor) stars only. In principle, the latter could be evaluated within the same formalism via Equation~\ref{eq:GRB.efficiency}, but this is beyond the scope of this paper. Calibrating to the all-sky rate of GRBs found by the {\em Swift} mission of $\sim 100-1000$ bursts per year \citep[][]{Gehrels2009}, the combined Pop~I/II GRB formation efficiency is of order $\eta_{\rm GRB}\sim 10^{-9}-10^{-8}\, M_{\odot}^{-1}$ \citep[e.g.][]{BrommLoeb2006, Jaacks2019}. Given its flux-limited sensitivity, the {\em Swift} sample of GRBs is not expected to contain any Pop~III GRBs, whose detection has to await the next generation of space-borne missions (see next section).

\section{Transient Observations}
\label{sec:observability}
It is conceptually useful to first assess the signature of Pop~III transients for a suite of idealized instruments of `perfect' sensitivity, where no such transient would be missed in a given survey campaign. This will allow us to derive pure probes of the primordial IMF, before addressing the capabilities of planned telescopes, whose limited sensitivity will imply the incomplete sampling of the high-redshift transient sources. In the latter case, we will suggest ways to correct for this incompleteness, and to focus on robust IMF probes that do not greatly depend on specific flux limits and other instrumental limitations.

\subsection{Using perfect telescopes}
\label{sec:obs.perferect}
Assuming perfect sensitivity, Figure~\ref{fig:2} depicts the {\rm absolute} rate of events for PISNe (top row) and GRBs (bottom row) from Pop~III sources at $z < 10$, derived from Equation~\ref{eq:fov.times} with the combined assumptions in Section~\ref{sec:model}. For PISNe, rates are given for a 10~$\deg^{2}$ FOV, comparable to planned transient surveys with the {\em RST}, whereas GRB rates are for the entire sky. Here, both transient source rates are shown as functions of IMF slope, ($\alpha$; left column), and of the average stellar mass ($\bar{m}$; right column). In addition, all curves are color-coded by the characteristic IMF mass, $m_{\rm char}$. The GRB rates are plotted using individual lines to clearly indicate the degeneracy in the IMF parameters based on a given GRB rate. Even for the idealized case of perfect telescopes considered here, Pop~III PISN events are expected to be quite scarce compared to the total rate of Type~II SNe in a comparable FOV, which is predicted to be larger by a factor of $\sim 10^{3}$ \citep{Mackey2003}. It is evident that for $\alpha \lesssim 0.5$ (or $\bar{m}\gtrsim 140\ M_{\odot}$), PISN number counts alone could not determine the IMF parameters with any precision, given the near-constant rates in this regime.

While not as abundant as PISNe, Pop~III GRB rates are, in principle, more discriminating in terms of IMF parameters (see Figure~\ref{fig:2}, bottom panels). Given the low all-sky rates for extended regions in the IMF parameter space, however, many years of surveying would be required (e.g. for $\alpha \lesssim 1.0$). Furthermore, the search for Pop~III GRBs is presenting a pronounced ``needle in a haystack'' challenge, as the total GRB rate from Pop~I/II dominates by a factor of $\sim 100 - 1000$, with Pop~III beginning to break even only at the highest redshifts \citep[e.g.,][]{Wang2012,Toma2016}. Current constraints based on the {\em Swift} GRB mission indicate $\sim 100-1000$ bursts per year over the entire sky \citep{Gehrels2009}, which were likely all from Pop~I/II progenitors. We again reiterate that the value and possible evolution of $\eta_{\rm GRB}$ for Pop~III sources is highly uncertain, implying similarly large uncertainties for the resulting GRB rates. As we will discuss next, we can significantly alleviate at least a part of the uncertainty, the one deriving from the modeling of the underlying Pop~III SFRD (see Section~2.3), by considering the ratios of different kinds of transients (here PISNe and GRBs). 

\begin{figure*}
    \centering
    \includegraphics[width=0.975\textwidth]{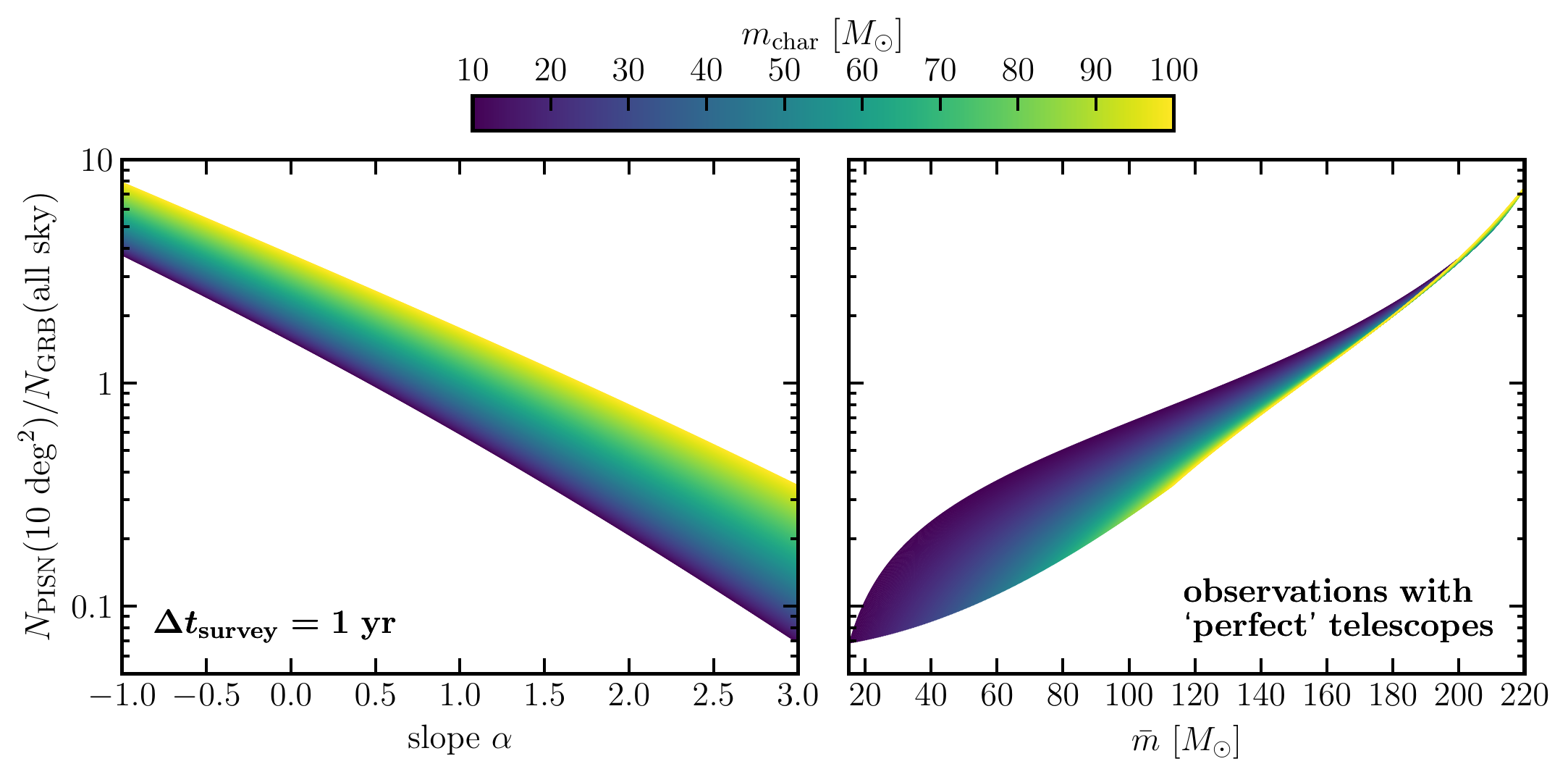}
    \caption{Constraints on the Pop~III IMF with combined transients. Curves are the ratios of the transient rates shown in Figure~\protect\ref{fig:2}, over the same observational time period, again as a function of the IMF slope ($\alpha$; left panel), and the average stellar mass ($\bar{m}$; right panel). As before, both panels assume ideal instruments without imposing any sensitivity limit. By combining the counts from both sources, we can effectively eliminate the degeneracies arising in the absolute counts considered separately. The transient ratio considered here, on the other hand, can in principle uniquely determine the shape of the IMF (see Section~\ref{sec:combined}). Since we assume the IMF not to evolve through time, this result should apply out to any redshift $z$.
    }
    \label{fig:3}
\end{figure*}

\subsubsection{Combining transient observations}
\label{sec:combined}
While absolute counts based on separate transient sources (PISNe and GRBs) could potentially probe the shape of the IMF, the previous section highlights the possible degeneracies arising from a given measured rate. We now wish to improve our ability to discriminate between different IMFs. When considering PISNe and GRBs in conjunction, we note that they behave very differently in terms of their respective duration, compared to a given survey time, $\Delta t_{\rm survey}$. GRBs are very short-lived, whereas PISN lightcurves have broad plateaus of a year or longer \citep[e.g][]{Kasen2011}, further prolonged through redshift effects. However, we assume that PISN sources can be monitored sufficiently long to establish their transient nature, which should be possible for $\Delta t_{\rm survey}\sim 1$~yr, based on existing modeling of the light curves (see below).

For a convenient comparison, we consider the (ideal) ratio of PISN sources observed over a FOV of $\Omega_{\rm PISN}$, for a given survey time, $\Delta t_{\rm survey}$ ($\simeq 1$ yr here), to the number of GRBs within $\Omega_{\rm GRB}$, observed over the same time:
\begin{align}
    \frac{N_{\rm PISN}}{N_{\rm GRB}}
    =  \left( \frac{\Omega_{\rm PISN}}{\Omega_{\rm GRB}} \right) \left(\frac{\eta_{\rm PISN}}{\eta_{\rm GRB}}\right) 
    \, .
\end{align}
Considering counts of PISNe over a 10~deg$^{2}$ FOV and GRBs over the entire sky, this ratio becomes
\begin{align}
    \frac{N_{\rm PISN}(10 \deg^{2})}{N_{\rm GRB}(\rm all\ sky)}
    \approx 2.4\times10^{-4}\, \left(\frac{\eta_{\rm PISN}}{\eta_{\rm GRB}}\right)
    \label{eq:ratio}
    \, .
\end{align}

We note that now, {\em we have avoided the considerable uncertainties in our knowledge of the Pop~III SFRD}, and have defined a much more robust diagnostic of the IMF. Implicitly, we here assume that the Pop~III IMF is not evolving in time. In reality, the transition from the predicted top-heavy Pop~III IMF to the observed bottom-heavy one for present-day stars may be gradual, introducing a possible redshift dependence \citep[e.g.][]{Clarke2003}. For simplicity, we will not pursue such a more complex scenario here. In Figure~\ref{fig:3}, we show the resulting ratio with the assumptions outlined above for $\Delta{t}_{\rm survey} = 1\ \rm yr$. The utility of this metric to robustly constrain the IMF parameters is evident, compared to the absolute counts of PISNe and GRBs alone. We again emphasize that this robustness is further enhanced by the (near-) redshift independence of this diagnostic. 

We further demonstrate this approach in Figure~\ref{fig:4}, where we present a ``worked example'' for a possible future experiment, again carried out in the context of our idealized perfect telescopes (we will relax this assumption later). Suppose that, by way of example, after multiple years of searching for Pop~III PISNe and GRBs, we were able to obtain enough statistics to infer a ratio of $N_{\rm PISN}/N_{\rm GRB} \simeq 1$ (within Poisson errors). Here, the intersection (pink) between the number of PISNe within a 10 $\deg^{2}$ FOV (magenta) and the all-sky GRB count (cyan) constrains the IMF in $\alpha-m_{\rm char}$ parameter space. Figure~\ref{fig:4} further emphasizes that relying only on individual absolute counts of either PISNe (magenta) or GRBs (cyan) can trace their respective regions in parameter space, but much tighter limits on the IMF shape can be obtained when they are combined (pink). Note that our specific choice here for the number of observed GRBs results in an extra band (bottom left), due to the degeneracy found for a fixed GRB rate within our model (see Figure~\ref{fig:2}). While not shown here, the same methodology could be applied to $\bar{m}-m_{\rm char}$ parameter space. Finally, we delineate the regions where the IMF would be top-heavy or bottom-heavy, separated by the vertical black-dashed line. While this example would not allow to unambiguously distinguish between top-heavy and bottom-heavy, a higher (or possibly much lower) value of $N_{\rm PISN}/N_{\rm GRB}$ would enable such distinction between these two regimes.

\subsection{Expectations for next-generation space telescopes}
\label{sec:obs.realistic}
For conceptual economy, we have carried out the previous discussion under the idealized assumption of telescopes with ``perfect'' sensitivity, such that even the faintest transients would be detectable. In reality, we are confronted with limited sensitivities and instrumental capabilities. Furthermore, this is a moving target, with technological progress re-defining such limitations. We will briefly discuss some of the next-generation telescopes currently planned for launch, or under development, to gauge the possibilities over the next decade or so. In the following subsection, we will further illustrate our overall methodology to translate observations from ``imperfect'' telescopes into the space of our idealized analysis framework.

Looking towards the future, the proposed THESEUS mission is designed to detect GRBs up to redshifts as high as $z\sim 20$ (see \citealt{Amati2018}). While the luminosity function for GRBs from Pop~III progenitors is an uncertain prior, we can assume that it follows a standard probability distribution \citep[e.g.][]{BrommLoeb2002}. Together with plausible assumptions on how the GRB rate is linked to the cosmic star formation rate, as a function of redshift, one can estimate that THESEUS will likely detect an order of magnitude more GRBs at all redshifts, with an even bigger efficiency gain at the highest redshifts (see \citealt{Ghirlanda2021}). Given its advanced capabilities, the THESEUS mission would be a close representation of our idealized `perfect' telescope, introduced previously. For THESEUS, or one of the other currently considered next-generation GRB missions, the remaining challenge would then be to distinguish a Pop~III GRB from the majority of Pop~I/II bursts, to be addressed further below. Significantly more daunting is the task to observe PISNe, which are both rare, and may not be bright enough for detection. The difficulty of uniquely distinguishing them from other SNe at high redshifts adds another aspect to the challenge (also discussed further below). Here, we will first assess whether a given PISN can be detected, in principle, by upcoming facilities using a back-of-the-envelope argument.

\begin{figure}
    \centering
    \includegraphics[width=\columnwidth]{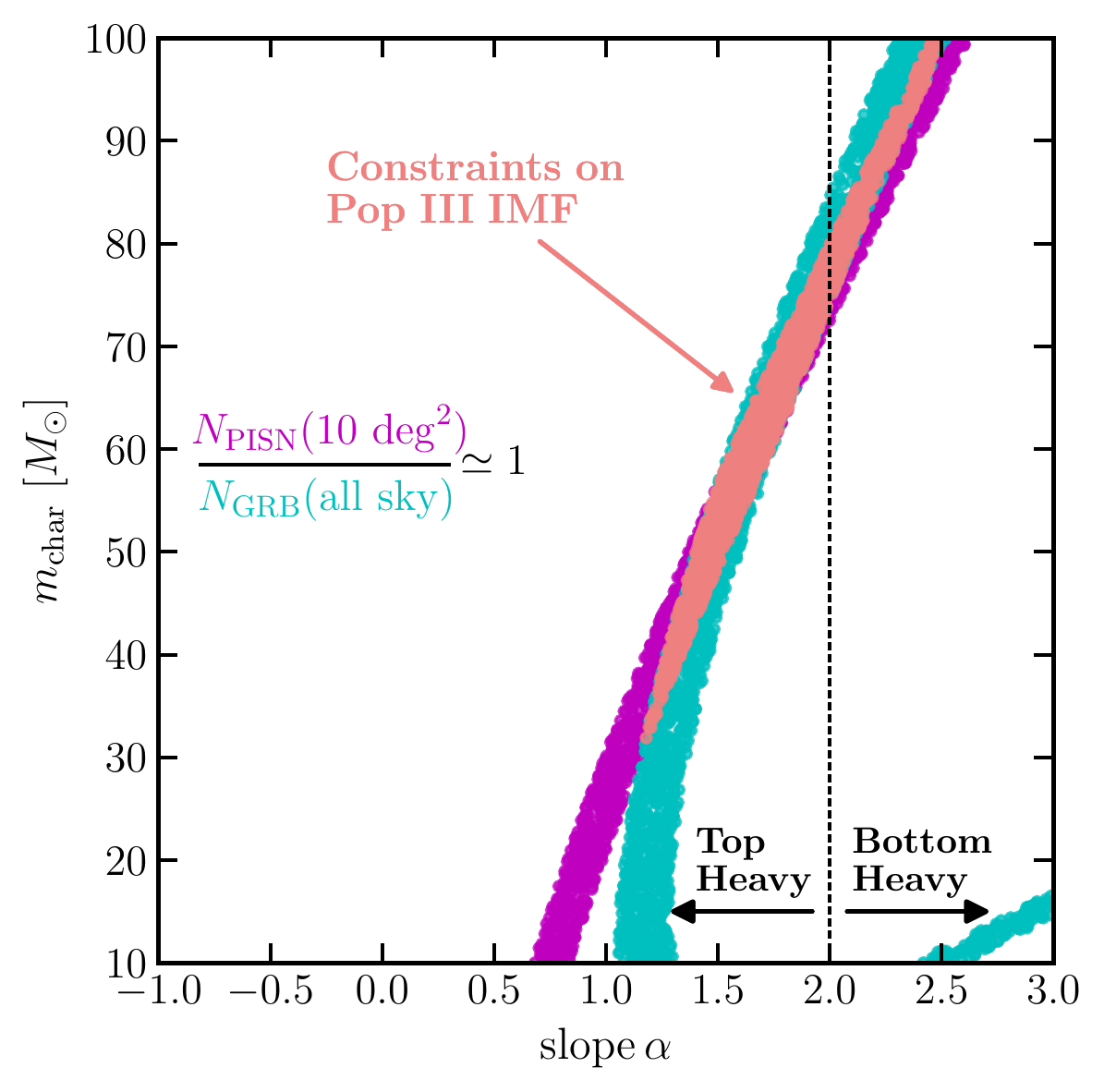}
    \caption{ Illustrative example for constraining the IMF in $\alpha$-$m_{\rm char}$ parameter space. We construct the possible outcome of a future experiment, again for the idealized assumption of perfect sensitivity, from the combined transient counts shown in Figure~\protect\ref{fig:3}. The number of PISNe (magenta) and the number of GRBs (cyan), over the same observation time, gives a ratio that can provide constraints for the IMF. It is clear that the combined constraint (intersection, highlighted in pink) performs better than the individual constraints.
    }
    \label{fig:4}
\end{figure}

A number of studies have considered the spectra and light curves of PISN explosions, to assess the prospect for their detection \citep[e.g.][]{Scannapieco2005,Weinmann2005,Frost2009,Kasen2011,Hummel2012,Pan2012,Souza2013,Whalen2013c,Souza2014,Hartwig2018b}. We here specifically consider the results from \cite{Hummel2012}, who in turn employed the light curves derived by \cite{Kasen2011}, with sophisticated radiation-hydrodynamical simulations.\footnote{Other simulations yield somewhat different predictions for the spectral time evolution \citep[e.g.][]{Dessert2013}, but a quantitative comparison of these different studies is beyond the scope of this paper.} A variety of PISN light curves are given in \cite{Hummel2012}; we bracket the models by considering the extreme cases of the faintest explosion (termed `B200') and the brightest one (`R250'), and we refer the reader to \cite{Kasen2011} for complete details. To approximately assess PISN visibility, we consider an efficiency factor to modify the counts for a perfect telescope, as follows:
\begin{align}
   \eta_{\rm vis}(z) := \frac{\Delta{t}_{\rm vis}(z)}{\widetilde{\Delta{t}}_{\rm PISN}}\, .
\end{align}
Here, $\widetilde{\Delta{t}}_{\rm PISN}\sim 10\ {\rm yr}$ is the approximate total duration of a PISN in the observer frame \citep{Hummel2012}, and $\Delta{t}_{\rm vis}(z)$ is the time, again in the observer frame, over which a given explosion at redshift $z$ would be detectable with a specific telescope, given its flux sensitivity. 

The upcoming {\em RST} has a long-wavelength limit of $\sim2$~microns \citep{Moriya2019}. Therefore, we consider only the F200W filter curves and, for the sake of simplicity, assume that the filters developed for {\em RST} will be comparable to the ones for NIRCAM on {\em JWST}. With {\em RST}, the R250 model is expected to be detectable out to $z\simeq10$ for its entire duration, such that $\Delta{t}_{\rm vis}\simeq10$~yr, resulting in a visibility factor of $\eta_{\rm vis}\approx 1$, while the B200 model with $\Delta{t}_{\rm vis}\simeq 0$ gives $\eta_{\rm vis}\approx0$. We conclude that realistic PISN detections will greatly differ from the idealized results (shown in Figures~\ref{fig:2} -- \ref{fig:4}). Therefore, even if most, or all, Pop~III GRB transients may be detectable with future missions such as THESEUS, the efficient detection of PISNe presents the operational bottleneck for our IMF diagnostic. With {\em RST}, we may be able to detect the brightest PISNe, but will likely miss the dimmer ones. Below, we will discuss an approach to still obtain useful IMF constraints, even if we were to miss many PISN sources in realistic surveys in the foreseeable future.

\subsection{Survey strategies}
In assessing the ability of next-generation telescopes and missions to constrain the Pop~III IMF, including the examples discussed in the previous subsection, we have to address these key challenges: How can we diagnose the Pop~III origin of a GRB observed at high redshifts? How can we distinguish a PISN from other explosions, such as superluminous SNe or hypernovae? Finally, how can we correct for the incompleteness in our Pop~III transient counts? The latter correction is crucial to map the observations into the idealized analysis space, as we have assumed perfect observational capabilities. We will briefly address these challenges in turn. 

\subsubsection{GRB identification}
\label{sec:GRB.id}
The proposed THESEUS mission promises to be sensitive enough to detect the prompt emission from even the highest-redshift GRBs, together with rapid $\sim$~arcmin localization. The latter will enable deep spectroscopic follow-up observations of the long-lived afterglow emission with {\it JWST} or the upcoming extremely-large ground-based telescopes \citep{Ghirlanda2021}. Indeed, the gas-rich circumburst environment in Pop~III star forming regions is predicted to result in bright afterglows, with peak observed wavelengths from near-IR to mm bands and fluxes of mJy to Jy, depending on the observed time after trigger \citep[e.g.][]{Wang2012}. Thus, to first order, these next-generation missions are designed to capture {\it all} GRBs throughout cosmic history, including the rare ones originating from Pop~III stars. 

For our purpose here, the focus therefore should be on disentangling the GRB origin, in terms of Pop~III or Pop~I/II. Identifying a unique identifier for Pop~III bursts has been a long-standing challenge \citep[reviewed in][]{Toma2016}. It is possible that the prompt emission of Pop~III GRBs is shifted to higher total explosion energies and/or longer durations, in response to the typically larger progenitor masses \citep{Meszaros2010}. A more robust diagnostic however is likely provided by the resulting afterglow. Pop~III afterglow spectra are expected to either show no metal lines, or weak ones originating in close vicinity to the burst, with primordial conditions farther away \citep{Wang2012}. One additional signature of a Pop~III burst may be the absence of any persistent emission from the underlying host galaxy, given their extremely low luminosities \citep[e.g.][]{Jaacks2019}. This is similar to the `no-host' class of dust-obscured, dark GRBs \citep[e.g.][]{Rossi2012}, but now the lack of any emission from the host would extend to all wavebands, not only the optical band in the classical case. However, Pop~II GRB hosts could be just as faint at comparable high redshifts \citep[see][]{Tanvir2012}.

\subsubsection{PISN identification}
For the PISN case, the situation is quite different. Here, the challenge is not to discriminate a Pop~III event from Pop~I/II, given that only Pop~III progenitors are expected to end up in the required PISN mass range (see above), but instead to distinguish the pair-instability case from CCSN events, including the rich class of super-luminous SNe \citep{Hartwig2018b}. Uniquely identifying a high-redshift PISN event is further complicated by the absence of any convincing observational template to date, rendering it a purely theoretical concept so far \citep{Barkat1967}. A few events have been discussed as PISN candidates, such as SN2006gy \citep{Smith2007} and SN2007bi \citep{Gal-Yam2009}, but the current consensus has moved to alternative explanations \citep{GalYam2012}. Indirect evidence for the existence of PISNe comes from stellar archaeology, with select metal-poor stars that may have preserved their peculiar nucleosynthetic yields \citep{Aoki2014}.

How then do we know that we have observed a PISN? One signature is the extended light curve, stretched by cosmological time dilation to possibly a decade in the observer frame for sources at $z\gtrsim10$. Furthermore, deep spectroscopic follow-up observations may be able to detect the progression from pure hydrogen/helium to metal lines, when the photosphere is receding deeper into the ejecta \citep[e.g.][]{Pan2012}. Finally, the same `no-host' argument for Pop~III progenitors applies as for GRBs (see above), reflecting the intrinsic faintness of any long-lived stellar component in the low-mass dark matter haloes of the first galaxies \citep[e.g.][]{Greif2007}.

\begin{figure}
    \centering
    \includegraphics[width=\columnwidth]{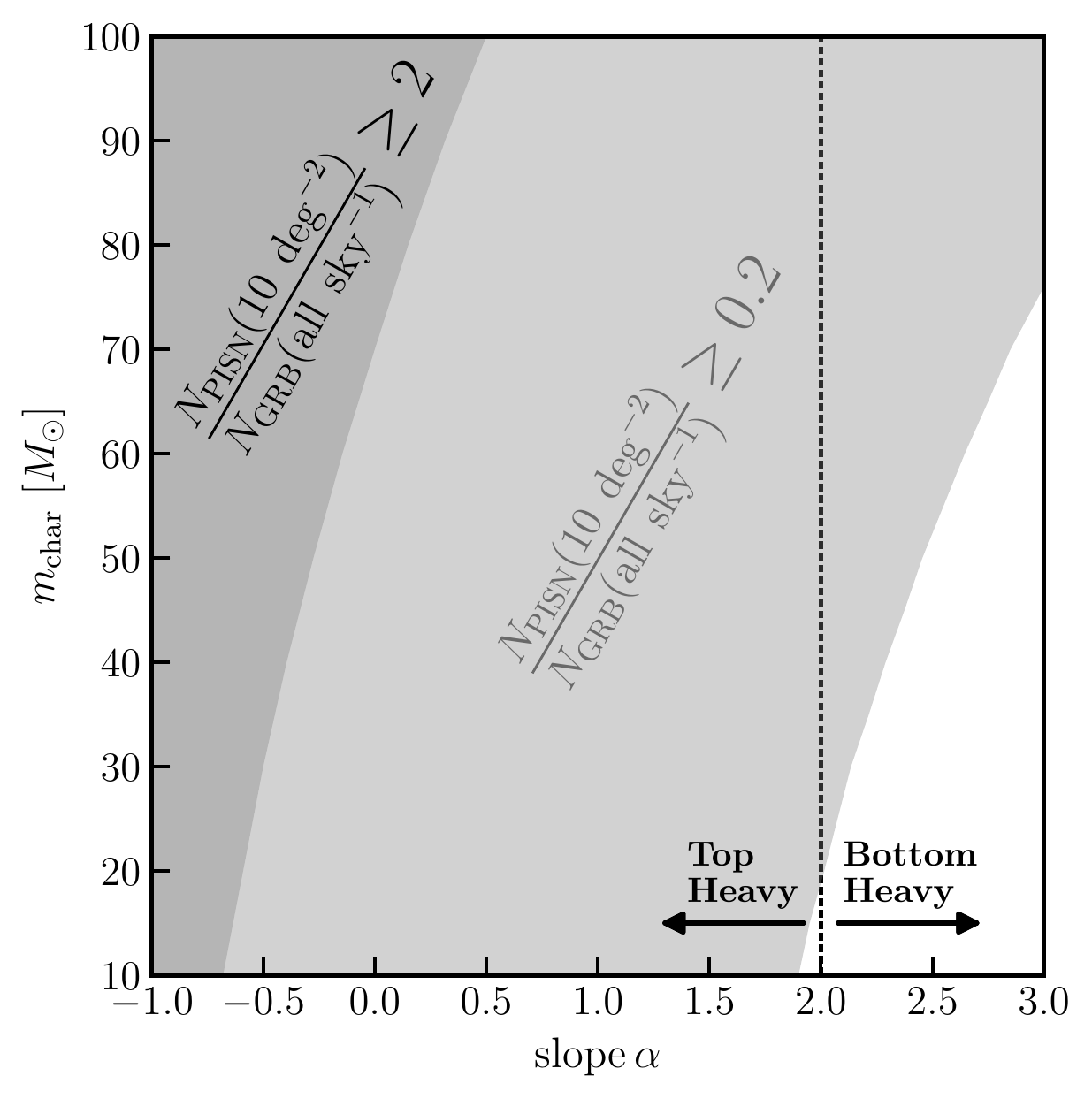}
    \caption{Illustrative example of lower bound constraints for the IMF. Similar in presentation to Figure~\protect\ref{fig:4}, but bands now encapsulate parameter space based on the lower limit of the number of observed PISNe. We here assume that the limiting factor is the number of observed Pop~III PISNe, with effectively {\it all} GRBs at high redshifts within reach of highly-sensitive, next-generation instruments (e.g. THESEUS).
    }
    \label{fig:5}
\end{figure}

\subsubsection{Missing transient sources}
With any combination of telescopes employed in the search for Pop~III GRBs and PISNe, we will have to contend with incomplete counts. Assuming that we are able to probe the Pop~III all-sky GRB rate perfectly, the key challenge will be to determine the PISN rate, given that we will likely miss a considerable fraction due to their faintness \citep[e.g.][]{Hummel2012}. However, even if we are unable to detect all Pop~III PISNe in a given FOV, as expected for a perfect instrument, {\em stringent bounds can still be placed on the Pop~III IMF parameters}. To illustrate this point, let us suppose that the $z\sim 9.4$ burst GRB~090429B \citep{Cucchiara2011} originated from a Pop~III progenitor star, observed in an all-sky survey over a period of about a year. To derive a similar, semi-empirical PISN rate, let us assume that the two archival PISN candidates inferred at redshifts $z\sim 2$ and $z\sim 4$ \citep{Cooke2012} had indeed Pop~III progenitors. Let us further assume that these PISNe resulted from a deep optical transient survey with an effective 10~$\deg^{2}$ FOV, observed for around a year, roughly in line with what can be inferred for the parameters of this archival search. Such lower-redshift PISN events, while rare, may in fact be possible, as several models predict that Pop~III star formation extends to redshifts as low as $z\approx 2$ \citep{Scannapieco2005,Tornatore2007,Trenti2009,Fumagalli2011,LiuBromm2020}. 

Combining the rough estimates from above would result in a lower limit of $N_{\rm PISN}(10\ \deg^{2})/N_{\rm GRB}(\rm all\ sky) \simeq 2$, based on our formalism. We show the corresponding region (dark gray) in $\alpha$ -- $m_{\rm char}$ parameter space in Figure~\ref{fig:5}. In this example, we would be able to determine that the IMF slope obeys $\alpha\lesssim -0.5$, with the strong implication that the IMF would be top-heavy, even though we would have almost no sensitivity on the characteristic mass. To illustrate this approach further, if future observations were to possibly find a smaller number of PISNe, or perhaps a larger number of GRBs, say, resulting in a factor of ten difference $N_{\rm PISN}/N_{\rm GRB} \simeq 0.2$, we would be unable to place stronger bounds on the shape of the IMF (light gray region in Fig.~\ref{fig:5}), even if a significant (and ultimately unknown) number of PISNe had been missed due to their faintness or confusion with other sources. Thus, imperfect, and possibly incomplete observations still offer the prospect of empirically constraining the vital character of the Pop~III IMF, which would be challenging to achieve with other indirect probes.

\subsection{Uncertainties and limitations}
In our modelling, we make a number of assumptions that are quite uncertain. We here focus on a few specific examples:

\begin{enumerate}
\item {\em SFRD}: We have calibrated our Pop~III SRFD to the peak value at $z\simeq 10$ from \cite{Jaacks2019}. While they are in broad agreement with simulation studies in the literature, model predictions of the Pop~III SFRD vary within one to two orders of magnitude. This would impact the inferences made from considering the absolute counts of either PISNe or GRBs (see Figure~\ref{fig:2}). However, working with the PISN to GRB ratio instead, the Pop~III IMF can effectively be probed regardless of any uncertainties in the underlying SFRD (see Section~\ref{sec:combined}).

\item {\em GRB efficiency}: We assume for simplicity that the formation efficiency of Pop~III GRBs does not depend on redshift. Moreover, our formulation assigns $\eta_{\rm GRB}$ assuming the statistical properties of Pop~III derived from numerical simulations, without any empirical calibration cross-check. Among the many uncertainties, if most Pop~III stars were not rapidly rotating, then we would expect that a significant fraction of the first stars would be unable to produce GRBs \citep{Stacy2011}. However, the work presented here provides a starting point for future improvements once better observational constraints on Pop~III star formation and high-redshift GRBs become available.

\item {\em Upper mass limit}: Our analysis assumes that Pop~III stars exhibit masses in the range of $m \in [1 - 300]\times M_{\odot}$, which is in line with the current understanding of the first stars \citep{Bromm2013b}. Of particular importance is the upper mass limit, as it is possible that the Pop~III IMF were limited to much smaller masses. Given that we have so far been unable to find any unambiguous hints of a Pop~III PISN, it is conceivable that the upper mass limit is much smaller ($m\lesssim 140 M_{\odot}$) than what is typically assumed. It would be straightforward to extend our analysis to such broader exploration of the upper mass limit, treated as an additional free IMF parameter. 
\end{enumerate}

\section{Summary and Conclusions}
\label{sec:conclusions}
Our understanding of the first stars has made great strides over the past decades. The detailed form of the Pop~III IMF, however, a key component of the emerging theoretical framework, remains highly uncertain. We here propose to employ high-redshift transients, specifically PISNe and GRBs, as effective probes of the Pop~III IMF. We are entering a new era of observational astronomy, where telescopes are becoming sensitive enough to detect the first galaxies, beyond the epoch of reionization ($z\geq 10$). Frontier missions, such as the {\em JWST} and {\em RST}, will greatly enhance our understanding of early cosmic history. The very first (Pop~III) stars will likely remain out of reach even then, given their intrinsic faintness as the smallest systems in the $\Lambda$CDM hierarchy. Alternative probes of the first stars are therefore essential, and we here demonstrate the promise of a comprehensive search for Pop~III transients. We find that PISNe and GRBs from Pop~III are rare out to redshift $z = 10$ (Figure~\ref{fig:2}), but an effective diagnostic of the Pop~III IMF can be constructed by considering these transients in conjunction, in particular alleviating parameter degeneracies, and the dependence on the underlying Pop~III SFRD. Importantly, this metric is able to provide tight constraints on the IMF even if source counts are subject to (not fully known) incompleteness corrections (Figure~\ref{fig:5}). 

The prospect for GRB cosmology in principle is bright, with a number of next-generation missions currently under development, such as THESEUS. Those missions can possibly probe all GRBs out to the highest redshifts ($z\sim 20$), including the still elusive Pop~III bursts. The hunt for PISNe will likely be more challenging. While the {\em RST} is expected to identify the brightest PISNe, it will still miss many of the dimmer transients, depending on the uncertain nature of the progenitor. Given all the uncertainties, it is promising that constraints on the IMF slope are still quite stringent, in particular, allowing us to answer the crucial first-order question of whether we are in a top-heavy regime or not. This basic character of the IMF in turn would determine how vigorously the first stars would have impacted the early Universe. The full power of the transient diagnostic could be harnessed when considered together with other, complementary probes of the Pop~III IMF, such as stellar-archaeology, which has already set conservative bounds on the IMF, for example on the lower mass limit \citep[e.g.][]{Hartwig2015}, and also constraints from the abundance of Pop III PISNe from the Cosmic Microwave Background \citep{Abe2021}. 

\section*{Acknowledgements}
We thank the anonymous referee for their helpful comments in improving the early version of this article. We thank Simon Glover for pointing out a numerical error in the early versions of this draft. AL was supported by the NASA grant 80NSSSC20K1469. The analysis in this paper made extensive use of the python packages {\small NumPy} \citep{Van2011}, {\small SciPy} \citep{Oliphant2007}, and {\small Matplotlib} \citep{Hunter2007}; we are thankful to the developers for these community tools. This research has made all intensive use of NASA's Astrophysics Data System (\url{https://ui.adsabs.harvard.edu/}), and the arXiv preprint service (\url{http://arxiv.org}).

\section*{Data Availability}
The data underlying this article is available in this article and the references therein.

\bibliography{references}

\bsp	
\label{lastpage}
\end{document}